\documentclass[twocolumn,tighten,times]{aastex61}

\usepackage{textcomp}

\newcommand{\rh}{$r_{\rm h}$~}
\newcommand{\Mgb}{\ensuremath{{\rm Mg}\, b}}
\newcommand{\Hbeta}{\ensuremath{{\rm H}\beta}}
\newcommand{\Feav}{\ensuremath{\langle {\rm Fe}\rangle}}
\newcommand{\Mgfep}{\ensuremath{[{\rm MgFe}]^{\prime}}}

\definecolor{purple}{rgb}{0.4,0.0,1.} 
\definecolor{lightgreen}{rgb}{0.67,0.87,0.0}

\accepted{by The Astrophysical Journal on March 19, 2018}

\shorttitle{Stellar populations of UCDs in M87}
\shortauthors{Zhang et al.}

\begin{document}

\title{Stellar population properties of Ultracompact Dwarfs in M87: a mass-metallicity correlation connecting low-metallicity globular clusters and compact ellipticals}

\correspondingauthor{Hong-Xin Zhang}
\email{zhhxin@gmail.com}

\author[0000-0003-1632-2541]{Hong-Xin Zhang}
\affiliation{Institute of Astrophysics, Pontificia Universidad Cat\'olica de Chile, Av.~Vicu\~na Mackenna 4860, 7820436 Macul, Santiago, Chile}
\affiliation{CAS Key Laboratory for Research in Galaxies and Cosmology, Department of Astronomy, University of Science and Technology of China, Hefei 230026, China}

\author[0000-0003-0350-7061]{Thomas H.~Puzia}
\affiliation{Institute of Astrophysics, Pontificia Universidad Cat\'olica de Chile, Av.~Vicu\~na Mackenna 4860, 7820436 Macul, Santiago, Chile}

\author{Eric W.~Peng}
\affiliation{Department of Astronomy, Peking University, Beijing 100871, China}
\affiliation{Kavli Institute for Astronomy and Astrophysics, Peking University, Beijing 100871, China}

\author{Chengze Liu}
\affiliation{Department of Astronomy, Shanghai Key Laboratory for Particle Physics and Cosmology, Shanghai Jiao Tong University, Shanghai 200240, China}

\author{Patrick C{\^o}t{\'e}}
\affiliation{NRC Herzberg Astronomy and Astrophysics Research Centre, 5071 West Saanich Road, Victoria, BC V9E 2E7, Canada}

\author{Laura Ferrarese}
\affiliation{NRC Herzberg Astronomy and Astrophysics Research Centre, 5071 West Saanich Road, Victoria, BC V9E 2E7, Canada}

\author{Pierre-Alain Duc}
\affiliation{Universit\'e de Strasbourg, CNRS, Observatoire astronomique de Strasbourg, UMR 7550, F-67000 Strasbourg, France}

\author{Paul Eigenthaler}
\affiliation{Institute of Astrophysics, Pontificia Universidad Cat\'olica de Chile, Av.~Vicu\~na Mackenna 4860, 7820436 Macul, Santiago, Chile}


\author{Sungsoon Lim}
\affiliation{Department of Astronomy, Peking University, Beijing 100871, China}
\affiliation{Kavli Institute for Astronomy and Astrophysics, Peking University, Beijing 100871, China}

\author{Ariane Lan\c{c}on}
\affiliation{Observatoire astronomique de Strasbourg, Universit\'e de Strasbourg, CNRS, UMR 7550, 11 rue de l'Universite, F-67000 Strasbourg, France}

\author{Roberto P.~Mu\~noz}
\affiliation{Institute of Astrophysics, Pontificia Universidad Cat\'olica de Chile, Av.~Vicu\~na Mackenna 4860, 7820436 Macul, Santiago, Chile}

\author{Joel Roediger}
\affiliation{NRC Herzberg Astronomy and Astrophysics, 5071 West Saanich Road, Victoria, BC V9E 2E7, Canada}

\author{Ruben S\'anchez-Janssen}
\affiliation{UK Astronomy Technology Centre, Royal Observatory Edinburgh, Blackford Hill, Edinburgh, EH9 3HJ, UK}

\author{Matthew A.~Taylor}
\affiliation{Gemini Observatory, Northern Operations Center, 670 North A'ohoku Place, Hilo, HI 96720, USA}

\author{Jincheng Yu}
\affiliation{National Astronomical Observatories, Chinese Academy of Sciences, Beijing 100012, China}


\begin{abstract}

We derive stellar population parameters for a representative sample of ultracompact dwarfs (UCDs) and a large sample of massive globular clusters (GCs) with stellar masses $\gtrsim$ 10$^{6}$  $M_{\odot}$ in the central galaxy M87 of the Virgo galaxy cluster, based on model fitting to the Lick-index measurements from both the literature and new observations.\ After necessary spectral stacking of the relatively faint objects in our initial sample of 40 UCDs and 118 GCs, we obtain 30 sets of Lick-index measurements for UCDs and 80 for GCs.\ The M87 UCDs have ages $\gtrsim$ 8 Gyr and [$\alpha$/Fe] $\simeq$ 0.4 dex, in agreement with previous studies based on smaller samples.\ The literature UCDs, located in lower-density environments than M87, extend to younger ages and smaller [$\alpha$/Fe] (at given metallicities) than M87 UCDs, resembling the environmental dependence of the Virgo dE nuclei.\ The UCDs exhibit a positive mass-metallicity relation (MZR), which flattens and connects compact ellipticals at stellar masses $\gtrsim$ 10$^{8}$ $M_{\odot}$.\ The Virgo dE nuclei largely follow the average MZR of UCDs, whereas most of the M87 GCs are offset towards higher metallicities for given stellar masses.\ The difference between the mass-metallicity distributions of UCDs and GCs may be qualitatively understood as a result of their different physical sizes at birth in a self-enrichment scenario or of galactic nuclear cluster star formation efficiency being relatively low in a tidal stripping scenario for UCD formation.\ 
The existing observations provide the necessary but not sufficient evidence for tidally stripped dE nuclei being the dominant contributors to the M87 UCDs.

\end{abstract}

\section{Introduction} \label{sec:intro}
In the size-luminosity plane, the division once thought to exist between globular clusters (GCs) and compact elliptical galaxies (cEs) has been blurred by the discovery of so-called ultracompact dwarfs \citep[UCDs;][]{hilker99, drinkwater00, phillipps01, hasegan05}.\ UCDs have been observationally defined \citep[e.g.][]{hilker09, brodie11} to be compact stellar systems (CSSs) with luminosities (10$^{6}$ $\lesssim$ $L_{V}$ $\lesssim$ 10$^{8}$ $L_{\odot}$) $\sim$ 0.5 -- 2.5 orders of magnitude higher than typical GCs and half-light radii (10 $\lesssim$ \rh $\lesssim$ 100 pc) which is at least several times larger than that of a typical GC.\ The intermediate nature of UCDs suggests that they may be either of galactic in origin (e.g.\ remnants of tidally disrupted nucleated galaxies; \citealt{bekki03, pfeffer13, pfeffer14}) or the scaled-up version of otherwise ``normal'' GCs (amalgamation of super star clusters (SSCs): \citealt{fellhauer02}; monolithic collapse of giant gaseous clumps: \citealt{murray09}).

It is non-trivial to differentiate between different formation mechanisms for UCDs, due partly to the lack of a complete theory for the formation of massive star clusters \citep[see][for some recent development]{kruijssen15, pfeffer18}, and partly to the utmost difficulty of detecting kinematical signatures of dark matter halos (if any) in CSSs \citep[e.g.][]{frank11}.\ Circumstantial evidence for the galactic origin of some UCDs include kinematical signatures of massive black holes \citep[e.g.][]{mieske13, seth14, ahn17}, signatures of tidal accretion events \citep[e.g.][]{norris11, jennings15, voggel16}, extended stellar envelopes \citep[e.g.][]{liu15}, and extended star formation histories \citep{norris15}.

The richness of both the UCD and GC systems appears to be most strongly correlated with the gravitational potential well in which the host galaxies reside \citep[e.g.][and references therein]{liu15, harris17}.\ The core regions of nearby rich galaxy clusters provide unique laboratories for systematically exploring the origins of UCDs with large and homogeneous samples \cite[e.g.][]{zhang15, liu15, voggel16, wittmann16}.~\cite{zhang15} carried out the first detailed study of the kinematical properties of the UCD system associated with M87 in the Virgo core region, and they found that the UCD system exhibits surface number density profiles, rotations and velocity dispersion anisotropies distinct from that of the GC system, suggesting that the two populations of compact stellar systems (CSSs) have either different formation mechanisms, different assembly histories, or/and different dynamical evolution histories.

In this paper, we present a comparative analysis of the stellar population parameters (including stellar mass, ages, metallicities and [$\alpha$/Fe] ratios) of a representative sample of M87 UCDs and GCs, in order to shed further light on the origin of the M87 UCD system.\ Throughout this work, we adopt the Virgo distance of 16.5 Mpc from \cite{blakeslee09} based on the surface-brightness fluctuation method.

\section{Data and Sample} \label{sec: datasample}
\begin{figure*}[t]
\centering
\includegraphics[width=0.95\linewidth]{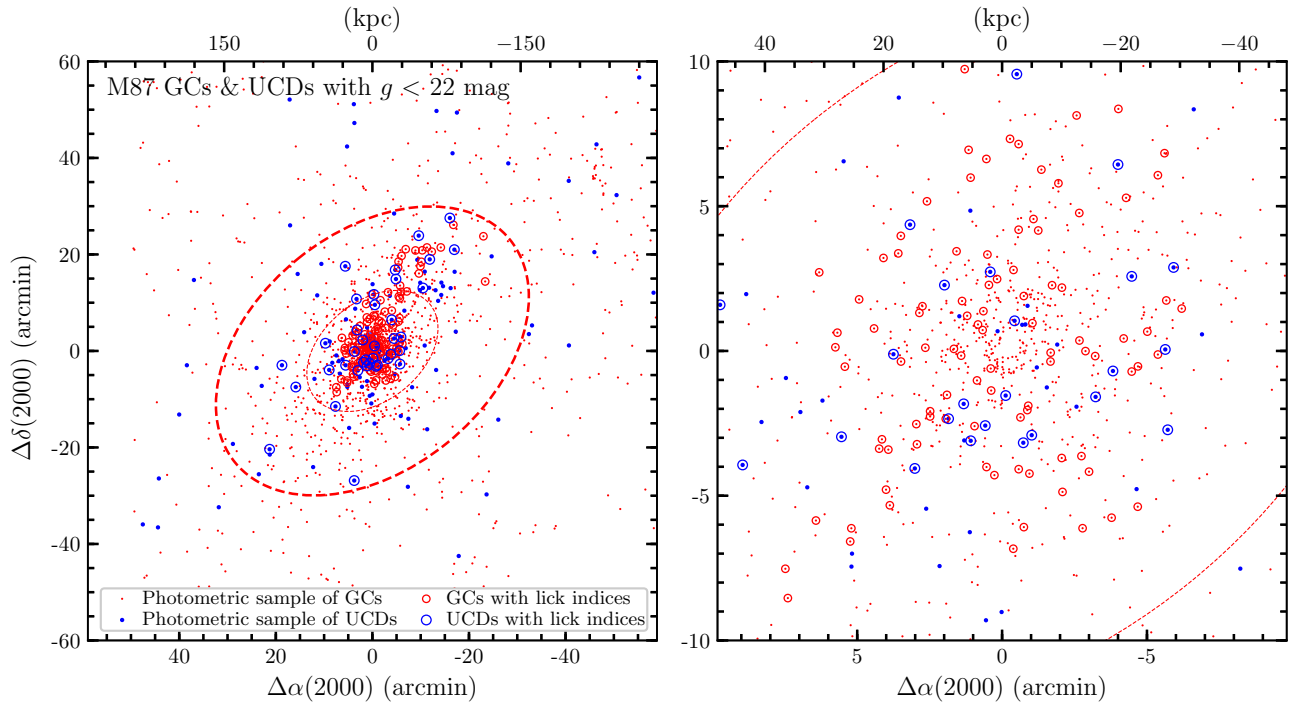}
\caption{Spatial distribution of M87 GCs and UCDs with $g$ $<$ 22 mag.\ North is to the top and East is to the left.\ The top axes represent the physical scales in kpc, for a distance of 16.5 Mpc.\ The center of M87 is at $\Delta\alpha(2000)$ = 0 and $\Delta\delta(2000)$ = 0.\ The {\it left} and {\it right} panels plot the central 120\arcmin$\times$120\arcmin~and 20\arcmin$\times$20\arcmin, respectively.\ The red and blue small dots, respectively, represent the photometrically selected GCs and UCDs.\ The red and blue circles, respectively, show the GCs and UCDs with Lick-index measurements.\ The thin dashed ellipse marks one effective radius of the number density profile of the UCD system, and the thick dashed ellipse marks a geometric average radius of 30\arcmin, which encloses the GCs and UCDs with Lick-index measurements.
\label{spatial_fullsample}}
\end{figure*}

\subsection{GCs and UCDs around M87}
The parent samples of GCs and UCDs around M87 are selected based on the optical $u^{*},g,r,i,z$ imaging data from the {\it Next Generation Virgo Cluster Survey} \citep[NGVS;][]{ferrarese12} and the near-IR $K_{s}$-band imaging data from the NGVS-IR project \citep{munoz14}.\ The broad available wavelength coverage, from $u^{*}$ to $K_{s}$, gives great leverage to efficiently distinguish the majority of Virgo CSSs and galaxies from the foreground stars and background galaxies \citep{munoz14}.\ In addition, the exquisite spatial resolution of the NGVS images (PSF FWHM $\sim$ 0.5\arcsec~in $i$ band) of the NGVS images allows us to separate the Virgo UCDs (\rh $\geq$ 10 pc) from GCs  (\rh $<$ 10 pc).\ Details about the selection of the UCD and GC samples are given, respectively, in \cite{liu15} and Peng et al.\ (in preparation).\ As in \cite{zhang15}, the full samples of GCs and UCDs are divided into ``blue'' and ``red'' subpopulations at ($g-i$) = 0.89 mag.

\subsubsection{Lick Indices of GCs and UCDs from the Literature}\label{ucdlick_literature}
\citet[][hereafter C98]{cohen98} presented measurements of the Lick indices\footnote{The corresponding paper lists the index values for H$\beta$, Mg{\it b}, Fe5270 and Fe5335.} of 150 M87 UCDs/GCs with apparent magnitudes $B\!<\!22.5$ mag based on deep optical spectra obtained from the LRIS instrument on the Keck {\sc I} Telescope.\ The C98 sample covers the central $\simeq$ 14\arcmin$\times$14\arcmin~around M87.\ Note that C98 did not separate UCDs from GCs in their analysis, due to a lack of the \rh information at that time.\ The sky coordinates of the C98 sample, as compiled by \cite{strader11a}, are matched with that of our parent sample in order to obtain the multi-band NGVS photometry and \rh measurements.\ By requiring a minimum S/N of 30 per 1.24 \AA~pixel, we end up with Lick-index measurements for 15 UCDs and 76 GCs from C98.

The measurement uncertainties of the Lick indices depend on the spectral S/N.\ We assign uncertainties $\sigma$ to the C98 Lick indices based on an exponential curve (i.e.\ $\log\sigma$ = A $\times$ B$^{-S\\/N}$ + C) fitting, for each index, to the tight relations between $\sigma$ and S/N at 5000 \AA~of our IMACS observations at similar resolution (see Section \ref{sec: newobs}).\ We used the four Galactic GCs observed both by C98 and \cite{puzia02} to derive the mean additive correction factors ($-$0.01 for H$\beta$, $-$0.18 for Mg$b$, $-$0.39 for Fe5270, and $-$0.16 for Fe5335) to calibrate the C98 Lick indices to the Lick/IDS system \citep{worthey97}.\ The standard deviations of the correction factors are 0.12, 0.12, 0.14, and 0.08, respectively, for H$\beta$, Mg$b$, Fe5270, and Fe5335.\

Besides the C98 sample, Lick indices of 12 of the most luminous M87 UCDs were measured by \citet[][hereafter E07]{evstigneeva07} with the Keck {\sc II} telescope, and \cite{firth09} and \cite{francis12} with the Gemini-North telescope.\ Of the 12 UCDs, one (Strom 417) is in common with the C98 sample.\ Among the three studies, only E07 calibrated their measurements for 7 UCDs to the Lick/IDS system through observations of Lick/IDS standard stars.\ Therefore, we choose to use the Lick indices determined by E07 for the 7 UCDs, and for the remaining 3 and 2 UCDs observed by \cite{firth09} and \cite{francis12} respectively, we calibrated the measurements to the Lick/IDS system using the mean additive correction factors (0.08 for H$\beta$, $-$0.06 for Mg$b$, $-$0.15 for Fe5270, and $-$0.05 for Fe5335) determined based on 3 objects in common with E07.\ Lastly, Lick-index-based stellar population parameter estimates of an additional UCD (S999) are presented by \cite{janz16}, the details of which will be given in Section \ref{sec: lickcss}.\ We will include S999 in our final sample of M87 UCDs.

\subsubsection{Lick indices of GCs and UCDs from new spectroscopic IMACS observations}\label{sec: newobs}
We obtained optical spectra of 18 UCDs and 51 GCs with $g$ $\leq$ 22 mag, using the IMACS multi-slit spectrograph on the 6.5-m Magellan Baade telescope in March 2016 (observing run CN2016A-58).\ The observations were made with the f/2 camera (FOV: $\sim $27\farcm5 $\times$ 27\farcm5), the 300 mm$^{-1}$ grism (1.341 \AA/pixel), and a slit width of 1\arcsec.\ The wavelength coverage is $\sim$ 3900--9000\AA.\ The spectral resolution is $\sim$ 6.5 \AA.

We used the photometric sample as input catalog for mask design and observed two masks, with one centered on M87 and the other one offset by 15\arcmin~to the NW along the major axis of M87 (see Fig.~\ref{spatial_fullsample}).\ The integration time was 3.5 hr per mask.\ We also observed 7 Lick/IDS standard stars of different spectral types (F9 to K1) for calibration purposes.\ The raw data was reduced with the COSMOS\footnote{The Carnegie Observatories System for MultiObject Spectroscopy} package.\ The spectral extraction and redshift measurement are respectively carried out using the IRAF {\sc apall} and {\sc fxcor} tasks.\ Finally, the spectra were degraded to the wavelength-dependent Lick/IDS resolution \citep[see e.g.][]{puzia13}, and then de-redshifted to the rest frame to be prepared for measuring the four Lick indices H$\beta$, Mg{\it b}, Fe5270 and Fe5335.

\begin{figure*}[t]
\centering
\includegraphics[width=0.93\linewidth]{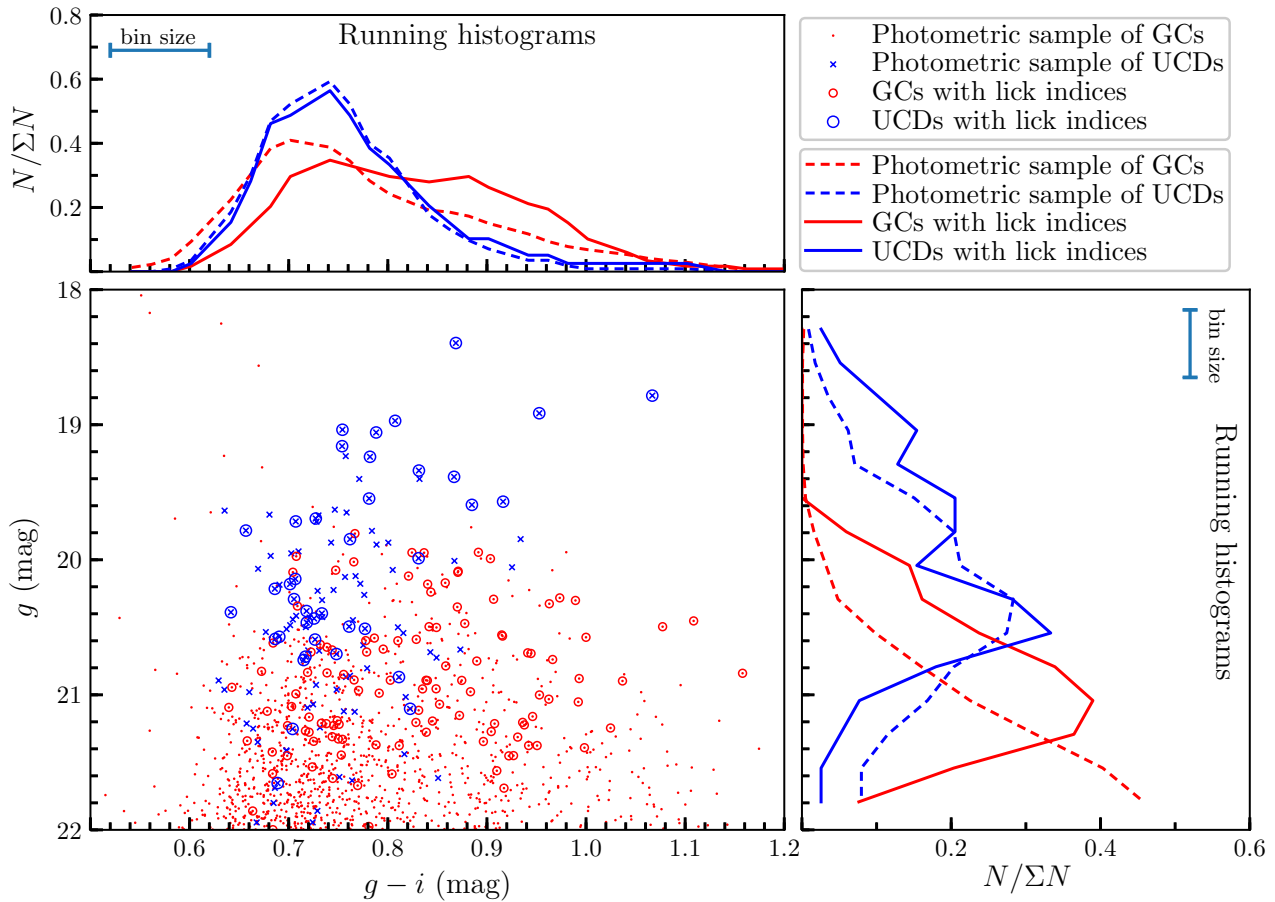}
\caption{
Color-magnitude distribution of the samples within the central 30\arcmin~of M87.\
The upper and lower right panels respectively show the running histograms of ($g-i$) and $g$.\
A bin size of 0.1 mag and a step size of 0.02 mag are used for the ($g-i$) running histograms.\
A bin size of 0.5 mag and a step size of 0.25 mag are used for the $g$ running histograms.\
The number counts in each running bin of the histogram have been normalized by the total number 
of objects in the corresponding subsample.\
\label{clrm_fullsample}}
\end{figure*}

The spectral S/N steadily decreases for fainter objects, with a S/N at 5000\AA~of $\simeq$ 30 pixel$^{-1}$ at $g$ $\simeq$ 19.5 mag and $\simeq$ 5 pixel$^{-1}$ at $g$ $\simeq$ 22 mag.\ For objects brighter than 19.5 mag, the Lick indices are directly measured on the individual spectrum.\ For objects fainter than 19.5 mag, the Lick indices are measured on the stacked spectra of the blue or red subpopulations in three $g$-mag bins divided at $g$ = 20 and 21 mag (see Fig.~\ref{clrm_fullsample}).\ The stacking is made over each Lick-index spectral window (encompassing both the feature and pseudo-continuum bandpasses) separately.\ For each index, the individual spectrum is first normalized by the mean flux of its pseudo-continuum before stacking.\ Each spectrum contributes equally to the stack.\ The noise spectrum for each stack is a quadrature sum of random uncertainties (propagated from the individual noise spectrum in the sample) and systematic uncertainties (from bootstrap resampling of the sample to be stacked).\ Uncertainties of the Lick indices are determined with Monte Carlo simulations, making use of the noise spectra.\ The Lick indices measured on the stacked or single GCs/UCDs are listed in Table \ref{lickindices_imacs16_tab}.

There is some overlap between our IMACS sample and the above mentioned literature sample (Table~\ref{lickindices_imacs16_tab}).\ We do not exclude these overlap objects from the stacking, in order not to introduce sample bias to the stack.\ Of the two brightest objects in our sample for which their individual spectra can be used for Lick-index measurement, VUCD3 is also in the E07 sample.\ The four Lick indices of VUCD3 measured by us and E07 are all in good agreement within 1-$\sigma$ uncertainties.


\subsubsection{Summary of the sample of GCs and UCDs}

Our final Lick-index sample includes 40 unique UCDs and 118 unique GCs, which, after the necessary spectral stacking, translates to 30 (28 individual and 2 stacked) sets of Lick-index measurements for UCDs and 80 (76 individual and 4 stacked) for GCs.\ The sample spans a $g$-mag range of 18 -- 22 mag and is confined within a maximum geometric average radius of $\sim$ 28\arcmin~from M87, equivalent to $\sim$ 2.2$\times$$R_{e, {\rm UCD}}$, where $R_{e, {\rm UCD}}$ is the effective radius of the bright UCD system \citep[at $g$ $<$ 20.5 mag;][]{zhang15}.\ The $g$-mag limit of 22 mag ($M_{g}$ $=$ $-9.09$) is 2 mag brighter than the turnover of the GC luminosity function (GCLF) of M87 \citep[e.g.][]{jordan07}, reaching the luminosity boundary between UCDs and GCs.\ Within the same magnitude limit and maximum radius, there are 143 UCDs and 1565 GCs in our photometric sample.\ The spatial distribution of the photometric and Lick-index samples at $g$ $<$ 22 mag is shown in Figure~\ref{spatial_fullsample}.

Figure \ref{clrm_fullsample} shows the ($g-i$) vs.~$g$ color-magnitude distribution of different subsamples.\ Normalized running-bin histograms of the colors and magnitudes are also shown.\ The Lick-index sample of UCDs is fairly representative of the corresponding photometric sample both in color and magnitude.\ The Lick-index sample of GCs has a disproportionally larger fraction of redder and brighter GCs than the photometric sample.

\begin{figure*}[t]
\centering
\includegraphics[width=\linewidth]{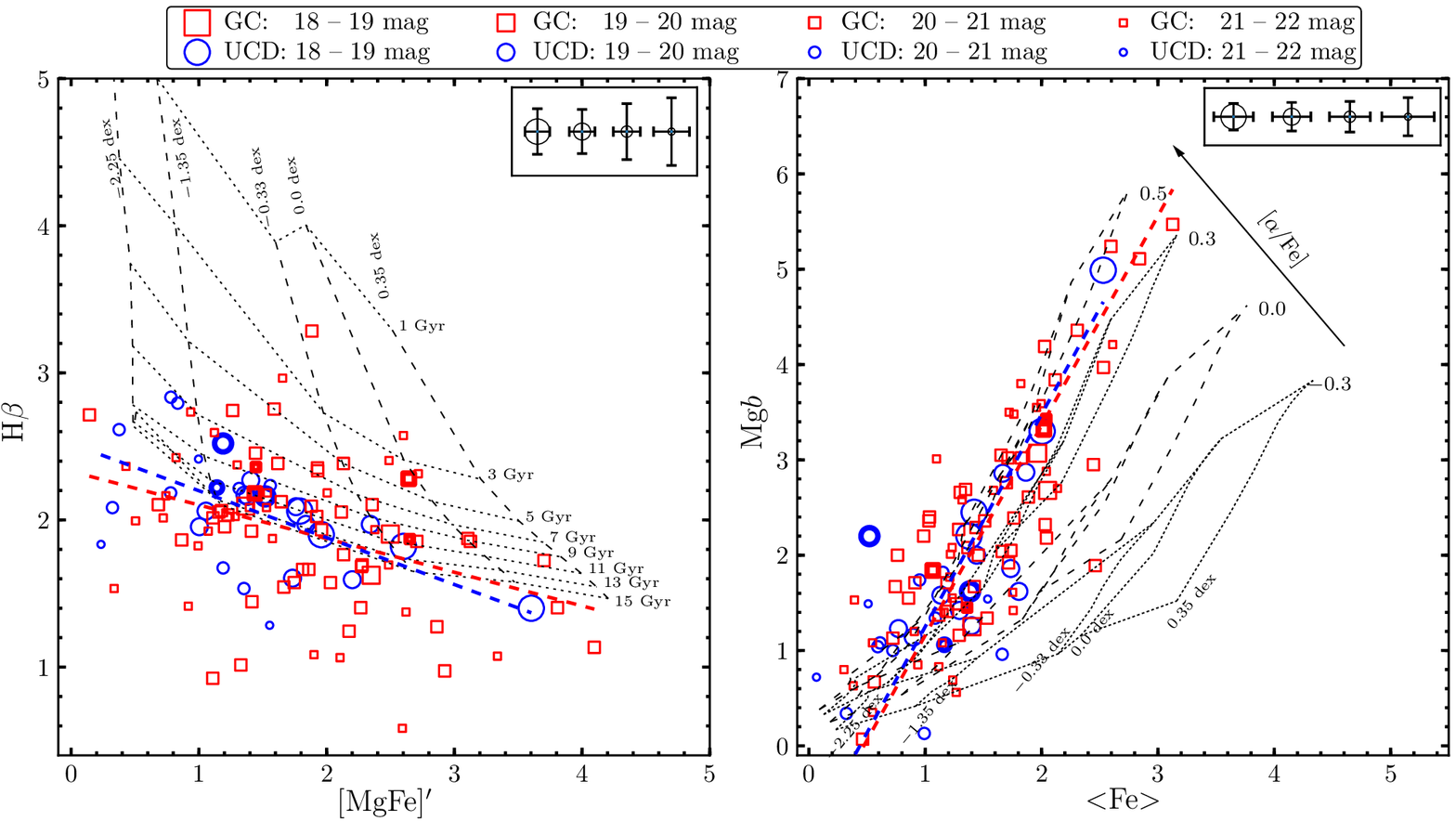}
\caption{Diagnostic plot distributions of M87 GCs ({\it red}) and UCDs ({\it blue}) in the \Hbeta$-$\Mgfep~and \Mgb$-$\Feav~planes.\ The thicker symbols mark the measurements based on the stacked spectra.\ The overplotted isochrone and iso-metallicity grid lines are from the SSP models of \cite{thomas11} with ages from 1 to 15 Gyr and metallicities [Z/H] from $-$2.25 to 0.35 dex.\ The model grids in the {\it left} panel are for a [$\alpha$/Fe] = 0.3.\ Objects that fall into different $g$-mag ranges are plotted with {\it circles} of different sizes.\ The median Lick-index errors of objects falling into different $g$-mag ranges are indicated in the upper part of each panel.\ The red and blue dashed lines mark the linear regression to the distributions of GCs and UCDs,  respectively.
\label{lickindex_licksample}}
\end{figure*}

\subsection{Lick indices of dE nuclei, cEs, and non-M87 UCDs from the literature}\label{sec: lickcss}
Besides the M87 GCs and UCDs, we also consider the literature Lick-index samples of the nuclei of Virgo dE galaxies, cEs and UCDs not belonging to M87, for comparison purposes.\ We restrict the selection of literature samples to those with either published Lick/IDS indices or stellar population parameters estimated with the same population model as adopted in this work for optimal consistency.

\citet[][hereafter P11]{paudel11} and \citet[][hereafter S17]{spengler17} respectively presented Lick/IDS index measurements for 26 and 19 dE nuclei in the Virgo cluster.\ To avoid potential biases of the Lick-index measurement at low spectral S/N, we opt to only include the dE nuclei with measurement uncertainties of the \Hbeta~and \Mgb~indices $\sigma_{{\rm H}\beta,{\rm Mg}b}\!<\!$ 0.5\,\AA\ and of Fe5270 and Fe5335 indices $<$ 0.6\,\AA, corresponding to a spectral S/N $\gtrsim$ 15 \AA$^{-1}$.\ With these selection criteria, we end up with 24 unique dE nuclei (20 from P11 and 4 from S17).

In addition, \cite{janz16} presented Lick-index-based stellar population modeling of a sample of 1 ultra-luminous GC (M85-HCC1), 1 M87 UCDs, 10 non-M87 UCDs and 17 cEs in the Virgo cluster and other environments.\ These objects have been discovered by various earlier studies \citep[see][for references]{janz16}.\ The 10 non-M87 UCDs include 4 (M59-UCD3, M60-UCD2, M60-UCD1, M59cO) in the Virgo cluster and 6 associated with S0 or ellipticals in group environments.\ \cite{janz16} did not publish their Lick-index measurements.\ However, \cite{janz16} used the same stellar population model of \citet[][hereafter T11]{thomas11} as in this work (see below) for fitting the Lick indices.\ So we can directly use the stellar population parameters published by \cite{janz16} in our comparative analysis.\ Lastly, Lick/IDS indices of 5 cEs in the Coma cluster as reported by \cite{price09} will also be included in our analysis.


\section{Analysis}
\subsection{Stellar population parameters}
We estimate the ages, metallicities [Z/H] and [$\alpha$/Fe] of M87 GCs and UCDs by fitting the T11 models to the measured Lick/IDS indices, following a procedure similar to that in \cite{puzia05} and \cite{graves08}.\ In particular, we adopt the \Hbeta$-$\Mgfep~and \Mgb$-$\Feav~diagnostic diagrams (Figure~\ref{lickindex_licksample}), where 
\begin{eqnarray*}
\Mgfep&=&\sqrt{\Mgb\,(0.72\!\times\!{\rm Fe5270}+0.28\!\times\!{\rm Fe5335})} \\
\Feav  &=&({\rm Fe}5250+{\rm Fe}5335)/2
\end{eqnarray*}
\citep[see][]{thomas03}, and iteratively invert the location of the measured indices on the two diagrams to stellar population parameters until a convergence is achieved.\ At each iteration, the ages and [Z/H] constrained by the \Hbeta$-$\Mgfep~diagram are used as input to constrain the [$\alpha$/Fe] through the \Mgb$-$\Feav~diagram.\ Because P11 and S17 adopted stellar population models that are different from our choice for fitting Lick/IDS indices, we choose to re-determine the stellar population parameters of their samples with the T11 model, in order to avoid model-dependent systematic biases in our comparative analysis.\ It is no doubt that using more Lick indices would offer stronger stellar population diagnostic power.\ Our choice of the four most commonly used Lick indices in the analysis is driven by the availability of such measurements in the literature.

The uncertainties of [Z/H], ages and [$\alpha$/Fe] are estimated by repeating the grid inversion for 150 realizations of the Lick indices of each object generated by randomly adding noise to the fiducial values according to their uncertainties.\ For objects that fall outside the model grid along the [Z/H] ($-$2.25\ to +0.67 dex) and [$\alpha$/Fe] ($-0.3$\ to +0.5 dex) dimensions, we do a linear extrapolation of the model grid.\ To those objects falling below the oldest iso-age grid of 15 Gyr \citep[see `\Hbeta~anomaly' in Fig.~\ref{lickindex_licksample}, see also][]{poole10}, we assign an age of 15 Gyr.\ We emphasize that an extrapolation of model grids is necessary mainly for [$\alpha$/Fe] but not for [Z/H] for our objects, so the following analysis in this paper will be mostly quantitative for [Z/H] but qualitative for [$\alpha$/Fe] and ages.\ The prime model-dependent uncertainties in our stellar population modeling are from the age estimates, due to the uncertain modeling of the \Hbeta~index.\ As we will present below, all of the UCDs and a vast majority of GCs have best-fit ages $\gtrsim$ 8 Gyr.\ A scatter of age estimates from 8 to 15 Gyr induces $<$ 0.2 dex uncertainty in the estimates of [$\alpha$/Fe] and [Z/H] (T11).

\subsection{Stellar masses}
Since the T11 models do not make predictions on the stellar mass-to-light ratio $\sc{M_{\star}}$/$L$, we opt to use the Flexible Stellar Population Synthesis (FSPS) SSP models \citep{conroy09} to estimate $\sc{M_{\star}}$/$L$ and thus $\sc{M_{\star}}$ of M87 GCs, UCDs, dE nuclei and the cEs from \cite{price09}, with the ages and [Z/H] from our Lick-index-modeling as input.\ For the non-M87 UCDs and the remaining cEs, we use the stellar mass as reported in \cite{janz16}.\ 

We note that several recent studies of extragalactic GCs found an apparent anti-correlation between the dynamical mass-to-light ratios and metallicities \citep[e.g.][]{strader11b}, which is  opposite to the predictions of the current stellar population synthesis models.\ Such anti-correlation, if real, may be driven by a metallicity-dependent variation of the stellar initial mass function (IMF).\ Nevertheless, more sophisticated modeling of the internal kinematics of MW GCs did not find such an anti-correlation, and instead found dynamical mass-to-light ratios in general agreement with those predicted by current stellar population models with a Kroupa or Chabrier IMF, with a possible exception for the metal-rich GCs \citep{baumgardt17}.\

\begin{figure*}
\centering
\includegraphics[width=0.98\linewidth]{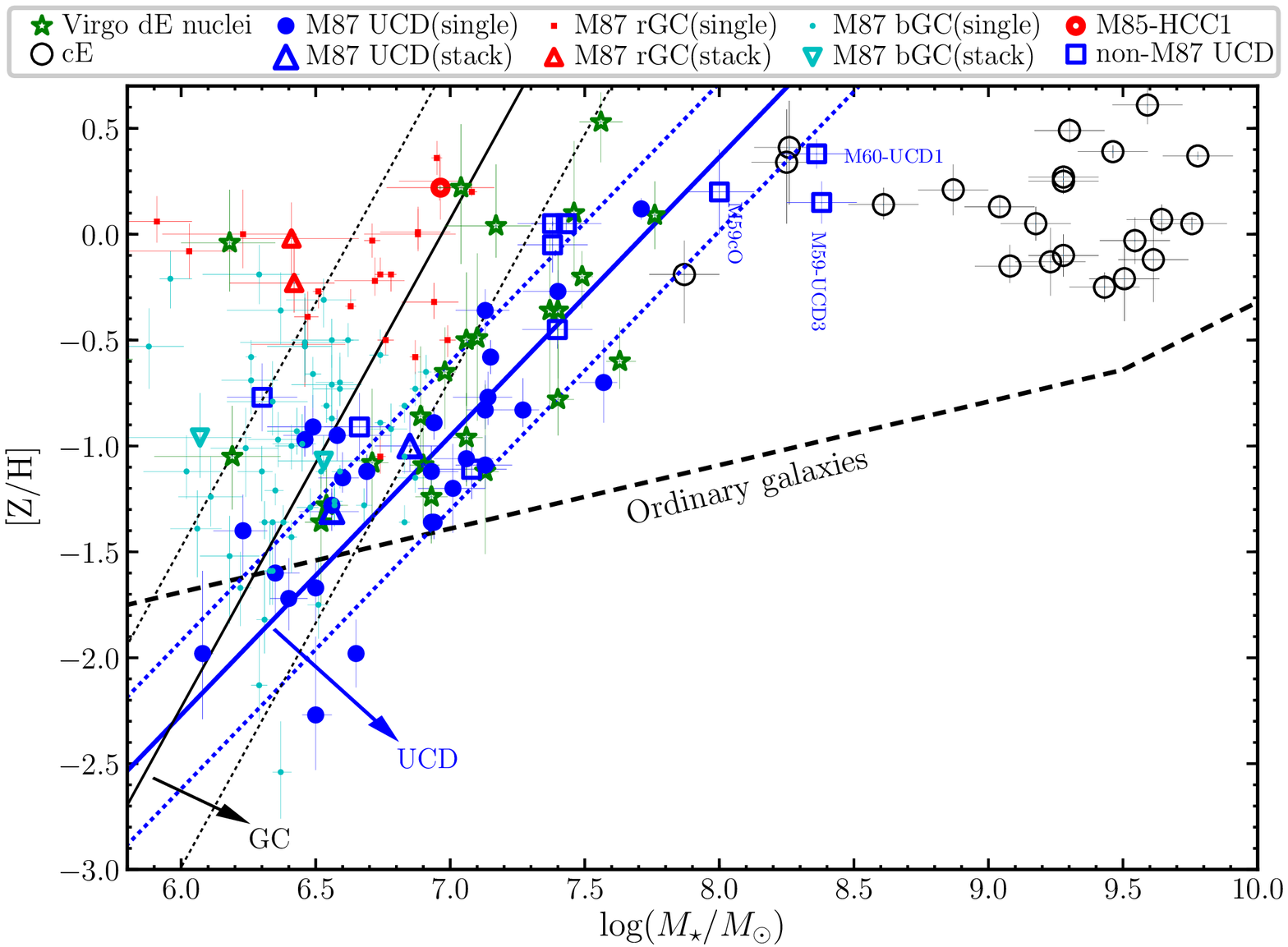}
\caption{Mass-metallicity distributions of the whole sample of CSSs.\ In the figure legend, ``rGC''  and ``bGC'' refers to the red and blue GCs, respectively.~M85-HCC1 is the densest known star cluster (\rh$\simeq$1.8 pc) discovered by Sandoval et al.\ (2015).\ The best-fit MZRs of M87 UCDs and GCs, as derived from a weighted ODR method, are marked as thick blue and thin black solid lines respectively, with dashed lines being the median 1-$\sigma$ scatter of [Z/H] around the relations.\ The best-fit parameters are given in Table \ref{mzr_tab}.\ The black dashed line marks the average MZR followed by dwarf galaxies with masses log($\sc{M_{\star}/M_{\odot}}$) $<$ 9.5 \citep{kirby13} and massive galaxies with log($\sc{M_{\star}/M_{\odot}}$) $>$ 9.5 \citep{gallazzi05}.
\label{stellpop_licksample_mzr}}
\end{figure*}

\begin{figure*}
\centering
\includegraphics[width=0.98\linewidth]{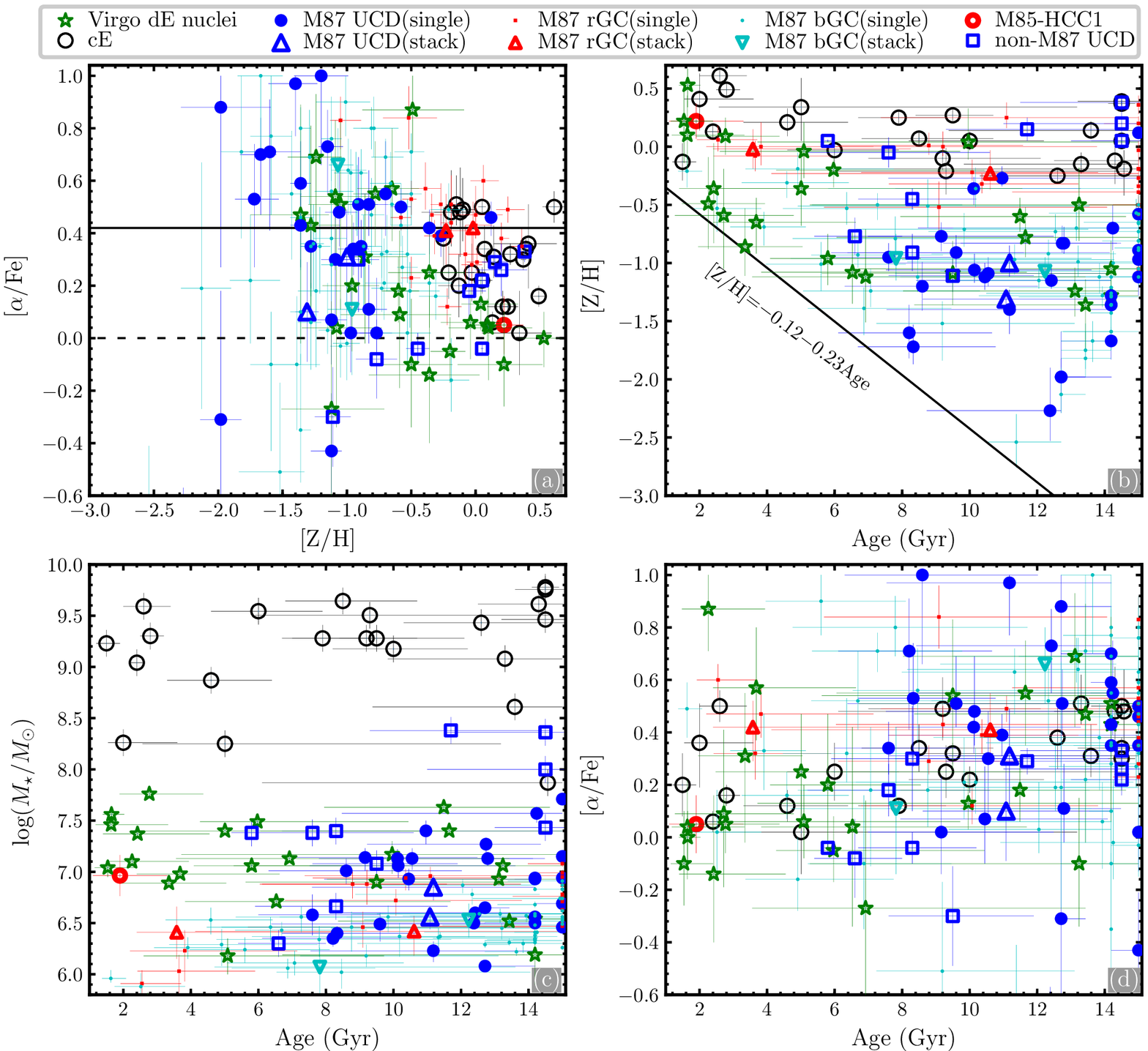}
\caption{As in Figure \ref{stellpop_licksample_mzr}, but here for distributions in various other diagrams involving stellar masses, ages, [Z/H], or [$\alpha$/Fe].\ In panel (a), a median [$\alpha$/Fe] = 0.42 for M87 GCs and UCDs is marked as a black solid line, while the dashed horizontal line indicates solar [$\alpha$/Fe] abundance ratios.\ In panel (b), the black solid lines mark the visually identified lower envelopes of the distributions of M87 GCs and UCDs.
\label{stellpop_licksample_others}}
\end{figure*}

\section{Results}
\subsection{Distributions of GCs and UCDs in the Lick/IDS index-index diagrams}\label{sec: indexindex}

The strength of the \Hbeta~and \Mgfep~indices respectively serve as the optimal age and [Z/H] indicators, whereas the index ratio \Mgb/\Feav~serves as a good indicator of the $\alpha$-element enhancement, especially at higher [Z/H] \citep{thomas03}.\ We have performed a weighted linear orthogonal distance regression (ODR) to the respective distributions of M87 GCs and UCDs in the two diagrams in Figure \ref{lickindex_licksample}.\ The lines of best-fit are over-plotted, and the best-fit equations are listed below.\
\begin{eqnarray*}
\Hbeta&=&-0.23\Mgfep + 2.33, \, \sigma=0.29 {\rm \AA}, \,\text{GCs} \\
\Hbeta &=& -0.32\Mgfep + 2.52, \,   \sigma=0.19 {\rm \AA}, \, \text{UCDs} \\
\Mgb &=& 2.20\Feav -1.04, \, \sigma=0.19 {\rm \AA}, \,\text{GCs} \\
\Mgb &=& 2.23\Feav - 0.98, \, \sigma=0.22 {\rm \AA}, \,\text{UCDs}
\end{eqnarray*}
The GCs and UCDs follow about the same relations on average.\ The standard deviation $\sigma$ of GCs with respect to the best-fit \Hbeta$-$\Mgfep~relation is larger than that of UCDs, which is either due to a larger age spread (see below) or a stronger influence of the H$\beta$ anomaly or uncertain horizontal branch (HB) morphology for GCs that are on average fainter than UCDs.~For the \Mgb$-$\Feav~distributions, GCs and UCDs have about the same $\sigma$ around their best-fit relations.

\subsection{Relations between [Z/H], [$\alpha$/Fe], $\sc{M_{\star}}$ and ages of CSSs}
Before proceeding to present our findings for Virgo UCDs and GCs and comparing with other types of CSSs, we briefly mention the relevant results from previous studies that are based on samples overlaping with ours.\ For Virgo UCDs, previous studies of a dozen of luminous UCDs (see references in Section \ref{ucdlick_literature}) found that UCDs generally have old ages ($\gtrsim$ 8 -- 10 Gyr) and super-solar [$\alpha$/Fe].\ For Virgo dE nuclei, \cite{paudel11} found a luminosity-metallicity correlation and a positive correlation between [$\alpha$/Fe] and local projected number density of galaxies.\ They also found that the dE nuclei in lower-density environments span a larger range of ages and [Z/H] that extends to younger and higher values.\ Regarding the comparison of the luminous Virgo UCDs and dE nuclei, \cite{paudel10} found that dE nuclei located in high density environment share similar stellar population properties (old ages and higher metal abundances) to UCDs.\ Lastly, \cite{janz16} noticed that, unlike lower-mass CSSs which have a large range of [Z/H], CSSs more massive than a few times 10$^{7}$ $M_{\odot}$ are exclusively metal-rich and deviate from the mass-metallicity relation (MZR) of ordinary galaxies towards higher metallicities at given stellar masses.\ 

Having collected a larger sample of Virgo UCDs and GCs with available spectroscopic stellar population parameters, we revisit the relationship between [Z/H], [$\alpha$/Fe], $\sc{M_{\star}}$ and age of different type of CSSs.\ The relevant diagrams are shown in Figures \ref{stellpop_licksample_mzr} and \ref{stellpop_licksample_others}.\ In the following subsections, we describe the most noteworthy trends shown in the figures.\

\subsubsection{$\sc{M_{\star}}$$-$[Z/H] relations}\label{sec:mzr}

We observe a positive mass-metallicity, i.e.~$\sc{M_{\star}}$$-$[Z/H] relation (MZR) for UCDs, with a Spearman's rank correlation coefficient $\rho$ of 0.76 for the M87 UCDs alone.\ The probability $p$ of the null hypothesis of no MZR is $9.9\times10^{-7}$.\ We fit a linear relation to the M87 UCDs with the weighted ODR method.\ The line of the best fit is over-plotted in Figure \ref{stellpop_licksample_mzr}, and the best-fit parameters are given in Table \ref{mzr_tab}.\ The standard deviation of [Z/H] around the best-fit relation is 0.35 dex, which is larger than the median of the measurement uncertainties of [Z/H] (0.12 dex).\ The MZR of UCDs extends up to log($\sc{M_{\star}/M_{\odot}}$) $\simeq$ 8.0 and [Z/H] $\simeq$ 0.2-0.3.\ At log($\sc{M_{\star}/M_{\odot}}$) $\gtrsim$ 8.0, the massive UCDs overlap with cEs on the $\sc{M_{\star}}$$-$[Z/H] plane, and the MZR flattens and ``saturates'' at [Z/H] of $\sim$ 0.2 dex, with a substantial scatter \citep[see also][]{janz16}.\

The Lick-index sample of GCs is not very representative of the photometric sample on the color-magnitude diagrams (e.g.\ Figure \ref{clrm_fullsample}).\ We alleviate the potential effect of this sample bias on the weighted MZR fitting of GCs by re-weighting the data points.\ In particular, we divide the ($g-i$) vs.~$g$ plane into 0.05$\times$1.0 mag cells, and then in each cell the data points are re-weighted by multiplying the measurement uncertainties by the square root of the number counts ratio of the Lick-index sample and the photometric sample.\ The best-fit MZR for the re-weighted sample of GCs is overplotted in Figure \ref{stellpop_licksample_mzr}, and the best-fit parameters of MZRs with/without re-weighting the data points are given in Table \ref{mzr_tab}.\ We note that the best-fit MZRs do not change significantly after re-weighting the data points.\

The GCs as a whole exhibit a much weaker mass-metallicity correlation than the UCDs (Table \ref{mzr_tab}).\ However, if only considering GCs with ages as old as UCDs, i.e.\ $\geq$ 8 Gyr (e.g.\ panel (b) of Figure \ref{stellpop_licksample_others}), there is a stronger and more significant MZR than the full sample of GCs.\ The GCs follow a steeper MZR than do the UCDs, and the {\it systematic} offset between the average MZRs of the two becomes larger at higher stellar masses.\ To further quantify the significance of the difference between the mass-metallicity distributions of GCs and UCDs, we perform a 2-dimensional KS test, following the numerical recipes described in \cite{press02}.\ The KS test suggests a 0.4\% (or 1.8\%) chance that the UCDs and GCs (or GCs $\geq$ 8 Gyr) are drawn from the same underlying distribution.\ The stronger MZR of the (M87) UCDs as compared to the GCs can be partly attributed to a larger range of stellar masses.\ In particular, if dividing the UCDs at log($\sc{M_{\star}/M_{\odot}}$)\ =\ 7.0 (i.e.\ the upper mass limit for GCs), neither the higher-mass nor the lower-mass UCDs exhibit a correlation with a significance level $p$ comparable to that of the full sample (Table \ref{mzr_tab}).\ 

All but three of the M87 UCDs in our sample are classified as blue UCDs, so we do not attempt to discuss the difference between the mass-metallicity distributions of the blue and red UCDs.\ For GCs, we present the best-fit MZR parameters for the blue and red subpopulations in Table \ref{mzr_tab}.\ The mass-metallicity correlations for the blue and red GCs are generally weak.

The Virgo dE nuclei appear to mostly follow the MZR established by UCDs, with $\rho$ = 0.55 and $p$ = 0.006.\ However, we point out that there may be a population of low-mass dwarf nuclei which did not pass our data quality selection function due to their low luminosities.\ Such objects may, in fact, have stellar masses and metallicities similar to those of red GCs.~Deeper spectroscopic observations are required to illuminate this aspect.

\subsubsection{[Z/H]$-$[$\alpha$/Fe] relations}

The Lick indices analyzed here provide weaker constraints on [$\alpha$/Fe] at lower [Z/H] (see Fig.~\ref{lickindex_licksample}).\ With this limitation in mind, we note that the M87 UCDs have overall higher [$\alpha$/Fe] at given [Z/H] (with substantial scatter) than the non-M87 UCDs, which are located in lower density environments.\ At [Z/H] $\lesssim$ $-0.5$, dE nuclei, M87 UCDs and blue GCs occupy very a similar parameter space in [$\alpha$/Fe] and [Z/H].\ However, at [Z/H] $\gtrsim$ $-$0.5, the Virgo dE nuclei generally have lower [$\alpha$/Fe], approaching solar abundance ratios, compared to the M87 UCDs, (red) GCs, and cEs.

\subsubsection{Trends with ages}
As a primary age indicator, the \Hbeta~absorption index is subject to various modeling uncertainties, such as the uncertain HB morphology, blue straggler stars, and stellar chromospheric emission fill-in from flaring stars \citep[e.g.][]{lee00, poole10}.\ Therefore, the age estimates presented here should be interpreted with caution.\ With these uncertainties in mind, we note that the youngest UCDs, GCs and dE nuclei are nearly exclusively the metal-rich ones, while those with older ages span a larger range of [Z/H] (panel (b) of Figure~\ref{stellpop_licksample_others}).\ In addition, the lower-mass GCs span a larger range of ages towards younger values (panel (c)), which, if real, might be explained if the recently accreted dwarf galaxies host a large fraction of relatively younger and lower-mass GCs \citep[e.g.][Ordenes-Brice\~no et al.~2018, in prep.]{usher15}.\ 

\section{Discussion and summary} 
\subsection{The mass-metallicity relation (MZR) of UCDs}
A tight optical color-magnitude relation for M87 UCDs was first noticed by \citep{brodie11}, who also found that the color-magnitude relation of UCDs is offset to bluer colors than that of blue GCs, but has a good coincidence with that of Virgo dE nuclei.\ Recent studies by \citep{liu15} with larger samples largely corroborated these findings.\ Here, we confirm that the color-magnitude relation of UCDs results from a remarkable MZR, and the offset from blue GCs or coincidence with dE nuclei is also explained by their different or similar {\it average} MZR.
 
The existence of a MZR might suggest a mass-dependent self-enrichment (e.g.~from supernova ejecta and stellar winds) within proto-cluster clouds.\ The deeper potential well of more massive systems means a higher retention efficiency of self-enriched gas that may be incorporated into the nearly simultaneous formation of low-mass stars or even the formation of subsequent stellar generations.\ This self-enrichment scenario \citep[e.g.][]{bailin09} has been invoked to explain the observed color-magnitude correlation, also known as the ``blue tilt'', of the blue GCs \citep[e.g.][]{strader06, mieske06, peng09}.

The observed MZR of UCDs appears to ``define'' the lower bound of the metallicity distribution of GCs at a given stellar mass (Figure~\ref{stellpop_licksample_mzr}).\ In a self-enrichment scenario, this might be simply explained by the larger physical size of UCDs.\ If the UCDs were born (through monolithic clump collapse) with larger half-mass radii, either due to larger sizes or shallower density profiles of the proto-cluster clouds, then at a given mass and star formation efficiency, the proto-cluster clouds of UCDs have lower gravitational potential energy \citep[i.e.~lower escape velocities, see also][]{janz16} and, thus, lower metal retention efficiencies than those of normal GCs.\ 

Alternatively, if UCDs were primarily formed as {\it dissipationless} mergers of SSCs or ordinary GCs, they would also be expected to be larger and less metal-rich than GCs (formed in monolithic collapsed clouds) of the same masses, because the merge increases the masses but not metallicities.\ Such merging can happen in extreme starburst environments \citep{fellhauer02} or nuclear regions of galaxies \citep[e.g.][]{lotz01, milosavljevic04, mclaughlin06, antonini15}.\ The latter possibility corresponds to the stellar cluster infall (driven by dynamical friction) scenario commonly invoked to interpret the formation of nuclear stellar clusters.\ Nevertheless, it is still to be seen how these dissipationless processes can result in a remarkable MZR rather than a wide spread of masses (or metallicities) for given metallicities (or masses).\ Lastly, if UCDs were initially formed as the stellar nuclei of (dwarf) galaxies and star formation efficiency in the nuclei was lower than similarly massive star clusters formed in isolation, they would also be expected to have lower metallicities than ordinary GCs for given stellar masses.\

\cite{pfeffer16} studied the formation of tidally stripped nuclei in the Fornax- and Virgo-like galaxy clusters using a semi-analytic galaxy formation model \citep{guo11}, and they predicted an average MZR for the stripped dE nuclei ($\log$[Fe/H] $\simeq$ $-$2.0 + 0.2$\log$$(M_{\star}/M_{\odot})$ for the Fornax and Virgo combined) that is much shallower than our observed MZR of M87 UCDs.\ This difference appears to be in disfavour with most UCDs (regardless of masses) being formed as tidally stripped dE nuclei.\ However, we note that the \cite{pfeffer16} model does not have a self-consistent treatment of the formation of nuclear clusters in galaxies of different masses and instead assigns metallicities to the stripped nuclei based on a fixed metallicity offset between the nuclei and their host galaxies.\ Therefore, it might still be premature to draw firm conclusions based on a comparison between the observed UCDs and the predicted stripped dE nuclei.

\subsection{Implications for the origin of UCDs}
The Virgo dE nuclei have ages ranging from $\sim$ 2 to 14 Gyr, whereas the M87 UCDs are almost exclusively $\gtrsim$ 8 Gyr, in agreement with previous studies based on smaller samples.\ Moreover, at [Z/H] $\gtrsim$ $-0.5$, the dE nuclei have smaller [$\alpha$/Fe] ratios and, thus, star-formation timescales longer than M87 UCDs (assuming equal IMFs and pre-enrichment).\ It is known that these young, metal-rich, and less $\alpha$-enhanced dE nuclei are almost exclusively found in relatively low-density environments \citep{paudel11, liu16}.\ Nevertheless, we find a similar environment-dependent difference between the M87 UCDs and non-M87 UCDs that are located in lower density regions, in the sense that the non-M87 UCDs have overall lower [$\alpha$/Fe] enhancements and extend to younger ages than the M87 UCDs.

Despite the above differences, the similar MZR of UCDs and dE nuclei implies a similar mass-dependent metal-enrichment history.\ The offset of the M87 blue GCs towards higher metallicities relative to this MZR is in line with an offset of blue GCs of other brightest cluster galaxies towards redder colors relative to the average color-magnitude relation of dE nuclei \citep{harris06}.\ While these observations do not necessarily mean that tidally stripped present-day dE nuclei are the primary contributors to the UCD population, they do suggest that the majority of GCs do not originate as stripped dE nuclei observed today.\ As discussed above, the difference between the mass-metallicity distribution of UCDs and normal GCs can be qualitatively understood as being simply due to a size difference, and (thus) the similar MZR of UCDs and dE nuclei could merely be a coincidence.

Above all, there is no significant difference between the stellar population parameters of UCDs and dE nuclei, as long as the environmental dependence is taken into account.\ This probably provides a {\it necessary} condition for favoring a galactic origin for most UCDs, but not a sufficient condition for ruling out that many UCDs (at least in M87) can be formed as the most massive and extended tails of the GC populations.\ In the former scenario, the system of UCDs is expected to have radially biased orbital structures, because only on highly radially biased orbits can nucleated dEs plunge deep into the central potential of the host galaxy in order to be tidally shredded to a naked nucleus.\ Such radially biased orbital anisotropies are indeed inferred for the M87 UCDs \citep{zhang15}.\ In the latter scenario, the UCD population may have been primarily formed {\it in situ} at early epochs or accreted from halos of earlier generations of dwarf galaxies which were characterized by larger average masses and more radially biased orbital structures than the more recently accreted dwarfs.\ The observed velocity field of blue GCs in M87 resembles that of the dE galaxies better than the UCDs, indicating that the blue GC population may have been continuously growing through accretion untill the present day.\ Whichever scenario is true for the Virgo core region, the present-day UCD system has been formed or/and assembled over a shorter timescale than the blue GC system.\

\begin{longrotatetable}
\begin{deluxetable*}{lccccccccccccccc}
\tabletypesize{\footnotesize}
\tablecolumns{16}
\setlength{\columnsep}{0.01pt}
\tablewidth{0pt}
\tablecaption{Lick indices of M87 UCDs and GCs measured based on Magellan/IMACS observations}
\tablehead{
\colhead{ID}
&\colhead{$N$}
&\colhead{$g$}
&\colhead{$\mu(g)$}
&\colhead{$(g\!-\!i)$}
&\colhead{$\mu(g\!-\!i)$}
&\colhead{H$\beta$}
&\colhead{Mg$b$}
&\colhead{Fe5270}
&\colhead{Fe5335}
&\colhead{SNR/pix}
&\colhead{Age}
&\colhead{[Z/H]}
&\colhead{[$\alpha$/Fe]}
&\colhead{$\log(M_{\star}/M_{\odot})$}
&\colhead{Comment} \\
\colhead{}
&\colhead{}
&\colhead{(mag)}
&\colhead{(mag)}
&\colhead{(mag)}
&\colhead{(mag)}
&\colhead{(\AA)}
&\colhead{(\AA)}
&\colhead{(\AA)}
&\colhead{(\AA)}
&\colhead{}
&\colhead{(Gyr)}
&\colhead{(dex)}
&\colhead{(dex)}
&\colhead{}
&\colhead{} \\
\colhead{(1)}
&\colhead{(2)}
&\colhead{(3)}
&\colhead{(4)}
&\colhead{(5)}
&\colhead{(6)}
&\colhead{(7)}
&\colhead{(8)}
&\colhead{(9)}
&\colhead{(10)}
&\colhead{(11)}
&\colhead{(12)}
&\colhead{(13)}
&\colhead{(14)}
&\colhead{(15)}
&\colhead{(16)} \\
\noalign{\vskip -4.5mm}
}

\startdata
\noalign{\vskip -0.5mm}
bGC\_stack\_g20.6     &  21   &  19.95$-$20.96  & 20.63  &  0.64$-$0.89  &  0.78 & 2.18$\pm$0.15 & 1.84$\pm$0.13 & 1.20$\pm$0.13 & 0.93$\pm$0.14  &  68.4 & 12.2$^{+2.1}_{-3.1}$ & $-$1.1$^{+0.1}_{-0.1}$ & 0.66$^{+0.14}_{-0.15}$ & 6.53$^{+0.16}_{-0.17}$ & 7 in C98 \\
bGC\_stack\_g21.4     &  15   &  21.19$-$22.00  & 21.42  &  0.66$-$0.83  &  0.72 & 2.36$\pm$0.34 & 1.45$\pm$0.29 & 1.55$\pm$0.30 & 1.17$\pm$0.32  &  31.4 & 7.8$^{+5.8}_{-3.1}$ & $-$0.96$^{+0.21}_{-0.19}$ & 0.11$^{+0.31}_{-0.25}$ & 6.07$^{+0.19}_{-0.27}$  & \nodata \\
rGC\_stack\_g20.7      & 4      &  20.29$-$20.90  & 20.69 & 0.91$-$1.04 & 0.94 & 2.28$\pm$0.33 & 3.32$\pm$0.28 & 2.21$\pm$0.30 & 1.83$\pm$0.30  & 31.5 & 3.6$^{+4.7}_{-1.2}$ & $-$0.02$^{+0.17}_{-0.17}$  & 0.42$^{+0.10}_{-0.13}$  & 6.41$^{+0.25}_{-0.26}$ & 1 in C98\\
rGC\_stack\_g21.3      & 11    &  21.16$-$21.45  & 21.31 & 0.90$-$1.02 & 0.93 & 1.87$\pm$0.37 & 3.43$\pm$0.32 & 2.06$\pm$0.32 & 2.02$\pm$0.31  &  29.1 & 10.6$^{+4.4}_{-6.5}$ & $-$0.23$^{+0.20}_{-0.14}$ & 0.41$^{+0.14}_{-0.13}$ & 6.42$^{+0.15}_{-0.24}$  & 1 in C98\\
bUCD\_stack\_g19.7   & 4      &  19.59$-$19.99  & 19.74 & 0.66$-$0.88 & 0.78 & 2.16$\pm$0.19 & 1.62$\pm$0.18 & 1.48$\pm$0.18 & 1.29$\pm$0.19 &  33.4 & 11.2$^{+3.0}_{-3.6}$ & $-$1.0$^{+0.15}_{-0.12}$ & 0.31$^{+0.18}_{-0.21}$ & 6.85$^{+0.11}_{-0.13}$ & 1 in C98\\
bUCD\_stack\_g20.4   & 12    &  20.14$-$20.87  & 20.42 & 0.64$-$0.81 & 0.71 & 2.22$\pm$0.18 & 1.06$\pm$0.16 & 1.31$\pm$0.17 & 1.02$\pm$0.17  & 56.4 & 11.1$^{+3.1}_{-2.5}$ & $-$1.31$^{+0.1}_{-0.1}$ & 0.1$^{+0.19}_{-0.19}$ & 6.56$^{+0.10}_{-0.12}$  & 3 in C98\\
H55930 (bUCD)   & 1      &  19.16$-$19.16  & 19.16 & 0.75$-$0.75 & 0.75 & 2.52$\pm$0.23 & 2.20$\pm$0.21 & 0.80$\pm$0.24 & 0.24$\pm$0.27  & 38.7 & 8.6$^{+3.4}_{-2.3}$ & $-$1.2$^{+0.2}_{-0.2}$ & 1.0$^{+0.0}_{-0.34}$ & 7.01$^{+0.12}_{-0.13}$  & \nodata\\
VUCD3 (rUCD)    & 1      & 18.78$-$18.78  & 18.78 & 1.07$-$1.07 & 1.07 & 1.22$\pm$0.17 & 5.18$\pm$0.14 & 2.66$\pm$0.15 & 2.37$\pm$0.15  & 61.8  & 15.0$^{+0.0}_{-0.1}$  & 0.12$^{+0.1}_{-0.0}$  & 0.48$^{+0.06}_{-0.06}$ & 7.71$^{+0.02}_{-0.03}$  & 1 in E07\\
\enddata
\tablecomments{
(1) ID.\ ``bGC'' and ``rGC'' respectively refers to blue and red GC, ``bUCD'' and ``rUCD'' respectively refers to blue and red UCD; 
(2) Number of objects used for the spectral stacking; 
(3) $g$-mag range; 
(4) median $g$ mag; 
(5) $g-i$ color range; 
(6) median $g-i$; 
(7-10) Lick/IDS indices;
(11) Signal-to-noise ratio at 5000 \AA~of the stacked or single spectra used for Lick-index measurement;
(12-14) Stellar population parameters derived by fitting with the Thomas et al.\ (2011) model;
(15) Stellar masses;
(16) Number of objects in common with \citet[][hereafter C98]{cohen98} or \citet[][hereafter E07]{evstigneeva07}.
}
\label{lickindices_imacs16_tab}

\end{deluxetable*}
\end{longrotatetable}

\begin{deluxetable}{lccccccc}
\tabletypesize{\footnotesize}
\tablecolumns{8}
\setlength{\columnsep}{0.01pt}
\tablewidth{0pt}
\tablecaption{Orthogonal Distance Regression to log[Z/H] = $a$ + $b$ $\times$ log($M_{\star}/M_{\odot}$)}
\tablehead{
\colhead{Subpopulation}
&\colhead{$a$}
&\colhead{$\sigma_{a}$}
&\colhead{$b$}
&\colhead{$\sigma_{b}$}
&\colhead{$\sigma_{[Z/H]}$}
&\colhead{$\rho$}
&\colhead{$p$} \\
\colhead{(1)}
&\colhead{(2)}
&\colhead{(3)}
&\colhead{(4)}
&\colhead{(5)}
&\colhead{(6)}
&\colhead{(7)}
&\colhead{(8)} \\
\noalign{\vskip -4.5mm}
}

\startdata
\noalign{\vskip -0.5mm}
UCD                        &  $-$10.17  &  1.07  &  1.32  &  0.15 &  0.35  &  0.76  &  9.9e-7 \\
UCD($>$10$^{7}$$M_\odot$)   &  $-$12.05  &  1.77  &  1.25  &  0.24 &  0.26  &  0.75  &  7.3e-3 \\
UCD($\leq$10$^{7}$$M_\odot$)    &  $-$2.89    &  1.95   &  1.25  &  0.29 &  0.41  &  0.41  &  8.2e-2 \\
GC                          &  $-$16.08  &  2.03  &  2.31  &  0.31  &  0.58  &  0.32  &  4.4e-3 \\
GC(rw)                    &  $-$17.61  &  2.47  &  2.56  &  0.38  &  0.76  &  0.32  &  4.4e-3 \\
GC($\geq$8Gyr)     &  $-$16.60  &  2.16  &  2.38  &  0.32  &  0.47  &  0.60  &  2.5e-7 \\
GC($\geq$8Gyr, rw)     &  $-$16.98  &  2.38  &  2.45  &  0.37  &  0.51  &  0.60  &  2.5e-7 \\
bGC                        &  $-$6.48  &  1.40  &  0.83  &  0.21  &  0.52  &  0.20  & 1.4e-1 \\
bGC($\geq$8Gyr)   &  $-$7.08  &  1.45  &  0.91  &  0.22  &  0.42  &  0.45  &  1.9e-3 \\
rGC                         &  $-$5.86  &  2.08  &  0.84  &  0.31  &  0.50  &  0.11  &  6.1e-1 \\
rGC($\geq$8Gyr)   &   $-$6.67  &  2.18  &  0.95  &  0.32  &  0.32  &  0.42  &  7.2e-2 \\
\enddata
\tablecomments{
The weighted orthogonal distance regression to the mass-metallicity distributions of different subpopulations in M87 are given in cols.\ 2-5 and the 
standard deviations of [Z/H] around the best-fit relations are given in col.\ 6.\ The spearman's ranking correlation coefficient $\rho$ and the two-sided 
$p$-values are given in cols.\ 7-8.\ GC(rw) and GC($\geq$8Gyr, rw) respectively refers to fitting to the samples of GC and GC($\geq$8Gyr) by re-weighting 
the data points in order to account for the biases of the Lick-index sample with respect to the photometric sample on the ($g-i$) vs.~$g$ color-magnitude diagram 
(see Section \ref{sec:mzr} for details).
}
\label{mzr_tab}

\end{deluxetable}

\acknowledgments
We thank the anonymous referee for his/her very constructive comments that lead to an improvement of the paper.\
This project is supported by FONDECYT Postdoctoral Grant No.~3160538, FONDECYT Regular Project Grant No.~1161817, and the BASAL Center for Astrophysics and Associated Technologies (PFB-06).\ HXZ also acknowledges a support from the CAS Pioneer Hundred Talents Program.\ C.L. acknowledges the NSFC grants 11673017, 11203017, 11433002.  C.L. is supported by Key Laboratory for Particle Physics, Astrophysics and Cosmology, Ministry of Education.\ EWP and SL acknowledge support from the National Natural Science Foundation of China through Grant No. 11573002.\ 

Observations reported here were partly obtained with MegaPrime/MegaCam, a joint project of CFHT and CEA/DAPNIA, at the Canada-France-Hawaii Telescope (CFHT) which is operated by the National Research Council (NRC) of Canada, the Institut National des Sciences de Univers of the Centre National de la Recherche Scientifique (CNRS) of France and the University of Hawaii.

\vspace{5mm}
\facilities{Magellan(IMACS), CFHT}

\end{document}